\documentclass[twocolumn]{aastex701}
\renewcommand{\textbf}[1]{#1}

\usepackage{graphicx}
\usepackage{txfonts}

\usepackage{xargs}                      
\usepackage{orcidlink}          
\usepackage{hyperref}
\hypersetup{
  colorlinks   = true, 
  urlcolor     = blue, 
  linkcolor    = blue, 
  citecolor    = blue  
}

\received{September 30, 2025}
\accepted{October 19, 2025}

\submitjournal{A\&A}

\begin{document}

\title{DKIST resolves sub-arcsec photospheric scattering polarization}

\author[0000-0002-3594-2247]{Franziska Zeuner}
\affiliation{Istituto ricerche solari Aldo e Cele Daccò (IRSOL), Faculty of Informatics, Università della Svizzera italiana, CH-6605 Locarno, Switzerland}
\email[show]{zeuner@irsol.ch}

\author[0000-0002-8775-0132]{Luca Belluzzi}
\affiliation{Istituto ricerche solari Aldo e Cele Daccò (IRSOL), Faculty of Informatics, Università della Svizzera italiana, CH-6605 Locarno, Switzerland}
\affiliation{Euler Institute, Faculty of Informatics, Università della Svizzera italiana, CH-6900 Lugano, Switzerland}
\email[noshow]{zeuner@irsol.ch}

\author[0000-0001-9095-9685]{Ernest Alsina Ballester}
\affiliation{Instituto de Astrof\'{i}sica de Canarias, E-38205 La Laguna, Tenerife, Spain}
\affiliation{Departamento de Astrof\'{i}sica, Facultad de F\'{i}sica, Universidad de La Laguna, E-38206 La Laguna, Tenerife, Spain}
\email[noshow]{zeuner@irsol.ch}

\author[0000-0001-6990-513X]{Roberto Casini}
\affiliation{High Altitude Observatory, National Center for Atmospheric Research, P. O. Box 3000, Boulder, CO 80307-3000, USA}
\email[noshow]{zeuner@irsol.ch}

\author[0000-0002-3215-7155]{David M. Harrington}
\affiliation{National Solar Observatory, Makawao, Hawaii, United States}
\email[noshow]{zeuner@irsol.ch}

\author[0000-0001-5131-4139]{Tanaus\'{u} del Pino Alem\'{a}n}
\affiliation{Instituto de Astrof\'{i}sica de Canarias, E-38205 La Laguna, Tenerife, Spain}
\affiliation{Departamento de Astrof\'{i}sica, Facultad de F\'{i}sica, Universidad de La Laguna, E-38206 La Laguna, Tenerife, Spain}
\email[noshow]{zeuner@irsol.ch}

\author[0000-0001-5131-4139]{Javier Trujillo Bueno}
\thanks{Affiliate scientist of the National Center for Atmospheric Research,\newline Boulder, U.S.A.}
\affiliation{Instituto de Astrof\'{i}sica de Canarias, E-38205 La Laguna, Tenerife, Spain}
\affiliation{Departamento de Astrof\'{i}sica, Facultad de F\'{i}sica, Universidad de La Laguna, E-38206 La Laguna, Tenerife, Spain}
\affiliation{Consejo Superior de Investigaciones Cientif\'{i}cas, Spain}
\email[noshow]{zeuner@irsol.ch}

\begin{abstract}
Scattering polarization signals offer a unique diagnostics of the physical conditions in the solar atmosphere, in particular magnetic fields via the Hanle effect. 
    However, their spatial structure remains poorly constrained due to the difficulty of achieving high spatial resolution and polarimetric sensitivity simultaneously. 
    We present the first direct observation of sub-arcsecond structuring in the linear scattering polarization of the photospheric Sr~{\sc i} 4607\,\AA\ line near the solar disk center ($\mu = 0.74$), obtained with the Visible Spectro-Polarimeter (ViSP) at the Daniel K. Inouye Solar Telescope (DKIST). 
    The data achieve about $\sim$0\farcs2 resolution with 30\,s integration and sufficient sensitivity to detect fine-scale patterns in the total linear polarization, which are evident in Sr~{\sc i} but absent in a nearby Fe~{\sc i} line that is simultaneously observed.
    Since this Fe~{\sc i} line is more Zeeman-sensitive than the Sr~{\sc i} 4607\,\AA\, this disparity  confirms that the signals in the Sr~{\sc i} 4607\,\AA\ line arise from scattering.
    These data provide the first spatially resolved two-dimensional maps of photospheric scattering polarization at sub-arcsecond scales, enabled by the capabilities of a 4-meter solar telescope.
\end{abstract}

\keywords{Polarization; Scattering; Methods: observational; Sun: photosphere; Techniques: polarimetric
               }

\section{Introduction}
\label{sec:intro}

Targeting small-scale magnetic fields in the solar photosphere is key to understanding the role of the Sun’s local dynamo in driving large-scale atmospheric dynamics.
The local dynamo is believed to generate complex and highly tangled components of this small-scale field, and to operate most efficiently in the highly turbulent regions of solar granulation--namely, the intergranular lanes \citep{Vogler2007,Rempel2014}.
Yet, direct observational constraints on these highly tangled fields remain elusive largely because they are invisible to standard Zeeman diagnostics. So far, some indirect evidence was provided by \citet[][]{TrellesArjona2021} using multiline inversions of intensity profiles, while \citet{TrujilloBueno2004} obtained more direct insights through modeling the scattering polarization of atomic and molecular lines.

Scattering polarization, modified by the Hanle effect, offers a unique window into this hidden magnetism \citep{Stenflo1982,TrujilloBueno2004}. 
Among the strongest and most promising signals for this kind of diagnostics is the one observed in the Sr~{\sc i} 4607\,\AA\ line (hereafter simply Sr~{\sc i}), long predicted to display spatial structure linked to small-scale atmospheric conditions, including magnetic fields.
However, directly resolving this spatial structure has remained an open observational challenge. 
Recently, statistical approaches have provided indirect evidence for it at disk center \citep{Zeuner2020,Zeuner2024}.

In this Letter, we report the first spatially resolved spectropolarimetric maps of Sr~{\sc i}, revealing scattering polarization in quiet Sun regions at sub-arcsec resolution.
These observations were obtained with the Visible Spectro-Polarimeter \citep[ViSP;][]{DeWijn2022} at the Daniel K. Inouye Solar Telescope \citep[DKIST;][]{Rimmele2022}.
By achieving sub-arcsecond resolution and polarimetric sensitivity better than $5 \times 10^{-3}$, our measurements confirm theoretical predictions and provide direct observational insight into the micro-structuring of scattering polarization in the lower solar atmosphere.
In \citet{Zeuner2025}, the authors present a detailed statistical analysis of Sr~{\sc i} scattering polarization across several limb distances, confirming the expected center-to-limb variation, quantifying the current sensitivity limits at disk center, and characterizing the signal distributions reported here.
These results demonstrate the capabilities of ViSP/DKIST and open a new path toward probing the solar small-scale magnetic field and evaluating the physical conditions that shape it.

\section{Observation and data processing}
\label{sec:data}

On September 1 2024, ViSP's second arm recorded a quiet-Sun region at $\mu = \cos(\theta) = 0.74$,\footnote{Product ID/Dataset ID: \href{ https://dkist.data.nso.edu/product/L1-WTXPV}{L1-WTXPV/BMNRV}} where $\theta$ is the heliocentric angle.
The spectrum covered a 8\,\AA\ spectral range centered on the Sr~{\sc i} 4607\,\AA\ line with a spectral sampling of 9.1\,m\AA\,pixel$^{-1}$. The data were obtained during the first ViSP operational cycle in which Sr~{\sc i} was offered, making these observations among the earliest of their kind.
A 61\farcs8 $\times$ 1\farcs1 field was scanned by stepping the slit in 20 steps over 10 minutes. 
We refer to each of these exposures at each step as a slit position, such that a sequence of slit positions forms a full scan.
Each slit position accumulates 115 modulation cycles with 10 modulation states. Each state is exposed for 12\,ms.
The scans are repeated two times over the same region.
We focus here on the scan with the best image quality, during which the adaptive optics (AO) system maintained full lock. 
The slit, with a width of 0\farcs0536, was oriented so that its position angle on the sky matched the parallactic angle. For the observed region, the slit tended to be parallel to the nearest solar limb.
Data were processed using ViSP’s standard reduction pipeline,\footnote{\url{https://docs.dkist.nso.edu/projects/visp/en/v2.16.7/l0_to_l1_visp.html}, which includes the DKIST system calibration \citep[see][]{Harrington2023}. } with the additional post-processing described in \citet{Zeuner2025}. 
Key steps are summarized below.

\begin{figure}[ht]
    \includegraphics[width=0.5\textwidth]{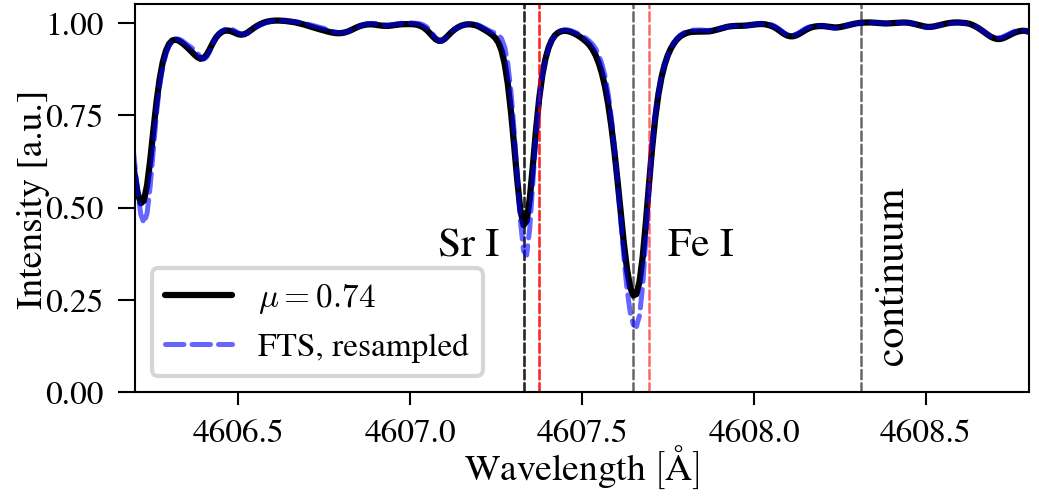}
    \caption{
    Intensity spectra from ViSP at $\mu=0.74$ compared to the FTS atlas \citep{Neckel1999}, resampled to match the ViSP spectral sampling.
    Dashed vertical black lines mark the line centers of Sr~{\sc i} and Fe~{\sc i}.
    Dashed red lines indicate the red-wing wavelength positions used for plotting Stokes $V/I$.
    All spectra have been normalized to the continuum for display purposes.
    }
    \label{fig:intensity_spectra}
\end{figure}

To reduce instrumental cross-talk from intensity into the polarized states, we applied the continuum-based part of the empirical correction by \citet{Sanchez1992}. We omit the full Mueller matrix calibration given the low polarization levels and suitable polarimetric accuracy of our data.
At $\mu = 0.74$, the assumption of negligible continuum polarization is justified \citep[well below 0.05\% at 4000\,\AA\ for $\mu > 0.7$;][]{TrujilloBueno2009}.
The continuum position used for this cross-talk correction is indicated in Fig.~\ref{fig:intensity_spectra}.
After this correction, we compute the fractional polarization by normalizing the Stokes parameters to the intensity at each pixel.
For simplicity, we refer to this as ``polarization'' throughout the text.

Owing to coordinate uncertainties ($\sim$7–10\arcsec), we forgo rotation of the polarization reference frame and instead present total linear polarization, $P_{\mathrm L} = \sqrt{(Q/I)^2 + (U/I)^2}$, computed at each wavelength.
Because absolute continuum polarization calibration is still being commissioned and improved, we subtracted the mean continuum across a 150\,m\AA\ window (to reduce statistical noise) in each polarization state, effectively forcing the continuum polarization to zero and avoiding misinterpretation. 
While some minor residual artifacts persist in the polarized spectra, their effect is controlled by consistently comparing Sr~{\sc i} with a neighboring Fe~{\sc i} line.\footnote{This line is a blend of two Fe~{\sc i} transitions.}

The spatially averaged intensity spectra show reasonable agreement with the straylight-free disk-center FTS atlas by \citet{Neckel1999} -- see Fig.~\ref{fig:intensity_spectra}, where the FTS data have been resampled to match the ViSP spectral sampling.
The mismatch in line profiles can be attributed to differences in disk position and to known internal ghosting in ViSP \citep[see][]{Zeuner2025}.

For both intensity and total linear polarization, we focus on two fixed wavelength positions: the line centers of Sr~{\sc i} and Fe~{\sc i}. 
These positions are marked by black dashed lines in Fig.~\ref{fig:intensity_spectra}.
Because circular polarization from the longitudinal Zeeman effect is weak in the line cores (in the absence of strong line-of-sight velocities), the circular polarization maps are obtained at the red wing wavelength indicated by red dashed lines in Fig.~\ref{fig:intensity_spectra}.

\section{Spatial resolution estimation and binning}
\label{sec:resolution}

\begin{figure}[h!t]
    \includegraphics[width=0.5\textwidth]{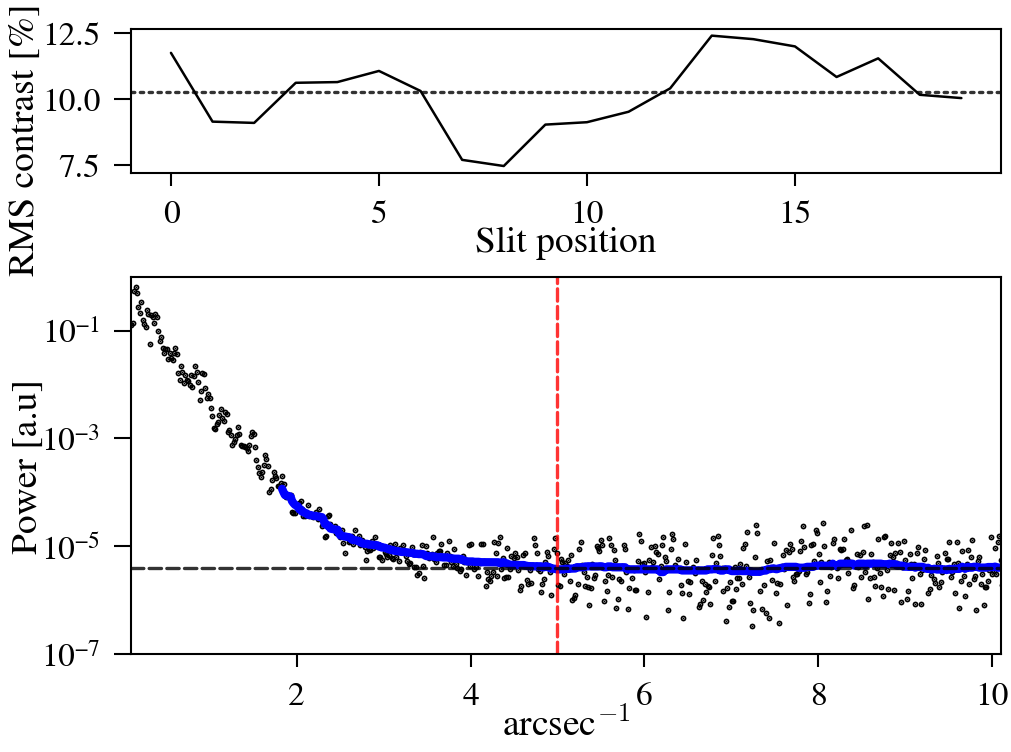}
    \caption{
        RMS intensity contrast variation with slit position (upper panel) and power spectrum of the intensity at the center of the Sr~{\sc i} line (lower panel). For the latter, the black dots show the power spectrum windowed by a Hann filter before computing the FFT, summed over all scan positions. The solid blue line shows a median filter applied to the Hann-windowed power spectrum. The dashed red vertical line indicates the frequency that equates to an effective spatial resolution of 0\farcs2.
    }
    \label{fig:bmnrv_power_srcore}
\end{figure}

To estimate the spatial resolution of our observation, we measured the apparent cut-off spatial frequency of the intensity power spectrum. 
We performed this analysis with the intensity map in the Sr~{\sc i} line center, but confirmed that using a continuum point yields very similar resolution estimates.
We first computed the root-mean-square (RMS\footnote{RMS contrast $=100\cdot
\frac{\sigma\!\left(I_{\mathrm{Sr}}\right)}{\langle I_{\mathrm{Sr}} \rangle}
$, where $\sigma$ is the standard deviation and $\langle \cdot \rangle$ is the spatial mean.}) intensity contrast as a function of slit position (see Fig.~\ref{fig:bmnrv_power_srcore}). 
Note that the RMS contrast is generally higher in the line core than in the continuum. 
We analyzed all slit positions to estimate a mean spatial resolution.
We applied a Hann window to each slit position to minimize edge effects that could introduce artificial high-frequency signals, before performing a one-dimensional fast Fourier transform (FFT). 
The resulting spatial power spectra reveal the distribution of spatial information across different spatial frequencies. 
By summing the power spectra from all slit positions, we enhance the signal-to-noise ratio, yielding a single, representative power spectrum shown in the bottom panel of Fig.~\ref{fig:bmnrv_power_srcore}.
Assuming the presence of additive white Gaussian noise, the flat high-frequency portion of the power spectrum provides an estimate of the noise floor (indicated by the black dashed line in Fig.~\ref{fig:bmnrv_power_srcore}). 
Signal power above this level is interpreted as real spatial information. 
Based on this analysis, we estimate an effective spatial resolution of approximately 0\farcs2 (red dashed line in Fig.~\ref{fig:bmnrv_power_srcore}).
It is important to note that the 30-second integration time per slit position likely reduces the effective spatial resolution compared to the theoretical limit set by spatial sampling. 
This degradation is due to solar evolution and residual image jitter.
The jitter has been estimated at the ViSP focal plane by an optical vibrometer to total around 25 to 35 milliarcseconds RMS variation \citep[see Figures 5 and 9 of][]{Sueoka2024} in DKIST data sets taken after 2022.
This leads to an order 0\farcs1 full-width-half-maximum image blur in long polarimetric exposures under reasonable atmospheric conditions and locked AO. 
Finally, we note that the spatial resolution may vary throughout the scan, depending on real-time atmospheric seeing conditions and the performance of the AO system.

To preserve spatial information while improving the signal-to-noise ratio, we spatially binned the scan to achieve a final spatial sampling of 0\farcs1 $\times$ 0\farcs1, which is sufficient for our analysis.

\section{Detecting scattering polarization in total linear polarization maps}

\begin{figure*}[ht]
\centering
\includegraphics[width=17cm]{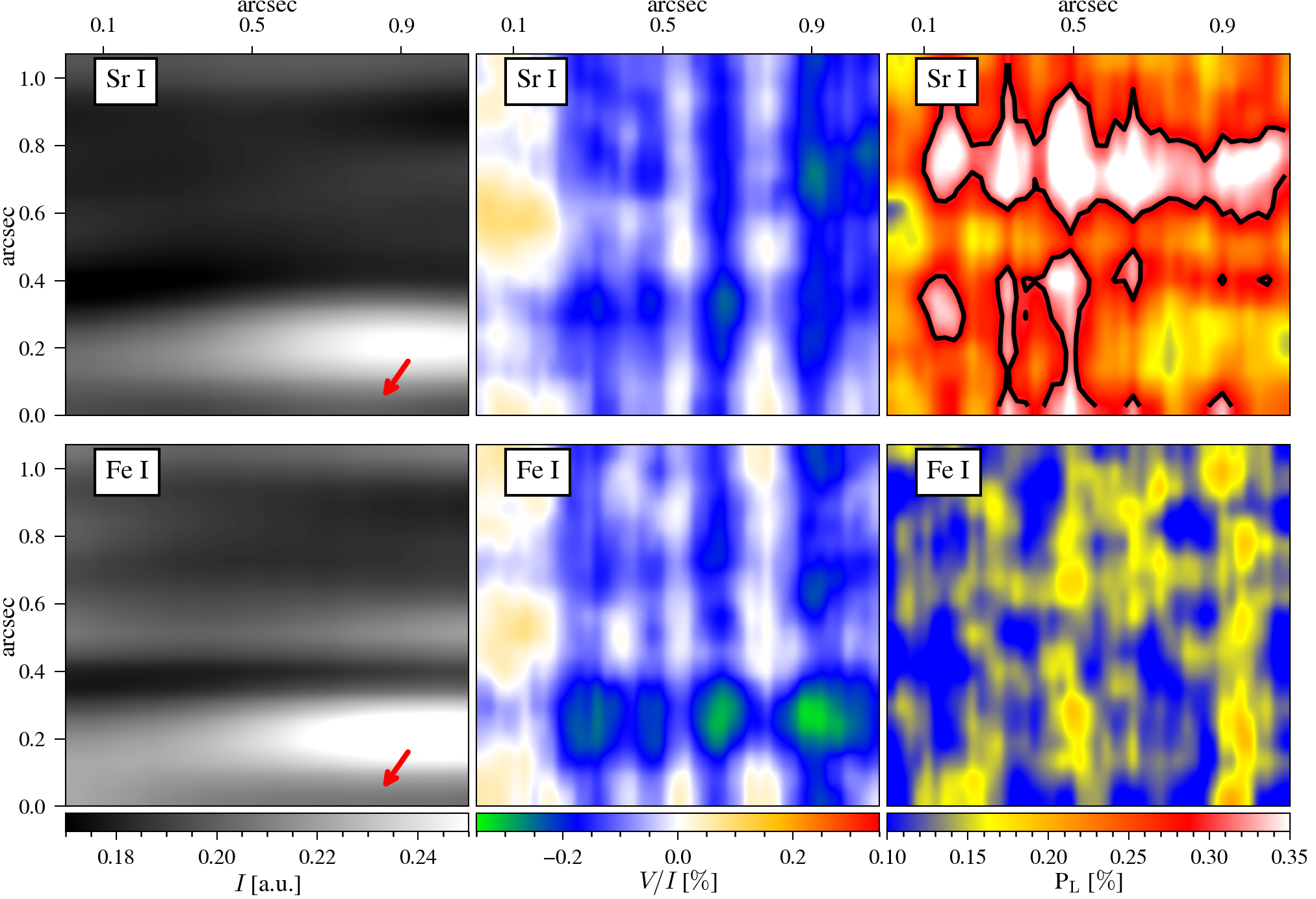}
    \caption{
       Intensity, circular polarization and total linear polarization maps (from left to right) in Sr~{\sc i} (top panels) and Fe~{\sc i} (bottom panels) of a 1\arcsec $\times$ 1\arcsec quiet Sun region scanned by ViSP at $\mu=0.74$. Data is spatially binned to 0\farcs1 $\times$ 0\farcs1 sampling. 
       Color bars for the polarization maps are applied consistently for both lines. 
       Note that the total linear polarization scale is saturated for Sr~{\sc i}.
       The approximate solar disk center direction is indicated by the red arrow in the intensity panels. 
       The scanning direction is upwards. 
       Contours in black are at the level of 0.3\%. 
       The intensity in Sr~{\sc i} is scaled by a factor of two to get the same intensity scale as for Fe~{\sc i}.
    }
    \label{fig:maps}
\end{figure*}

Figure~\ref{fig:maps} presents the maps of intensity, circular polarization ($V/I$), and total linear polarization ($P_\mathrm{L}$) for the Sr~{\sc i} and Fe~{\sc i} lines, extracted from a 1\arcsec $\times$ 1\arcsec sub-region of the full ViSP quiet Sun scan, treated as described in the last section.

Accounting for both photon and systematic noise, we conservatively estimate $\sigma_{\mathrm{Sr}}\approx0.1$\% and $\sigma_{\mathrm{Fe}}\approx0.14$\%, with the higher value for Fe~{\sc i} attributed to its deeper absorption profile, which reduces photon flux and thus increases the noise \citep[see][for details]{Zeuner2025}.

The selected sub-region shows exceptionally strong total linear polarization in Sr~{\sc i} with a maximum of 0.42\%, with signal levels well above the estimated noise.
White patches in the $P_{\mathrm{L}}$ map, defined by values $\geq 0.35$\%, exceed more than three times the noise level ($\sigma_{\mathrm{Sr}}$), corresponding to a signal-to-noise ratio (SNR) $\geq 3.5$.
Contours at $\mathrm{SNR}=3$ are overplotted to guide the eye.
These high-SNR signals are spatially structured on sub-arcsecond scales and appear both along the slit and in the scanning direction.
This spatial coherence across multiple slit positions rules out scan-induced or seeing-induced artifacts and strongly supports a solar origin.

A striking feature is the contrast in $P_\mathrm{L}$ between Sr~{\sc i} and Fe~{\sc i}: Sr~{\sc i} exhibits much stronger total linear polarization, although both lines show comparable spatial structuring in intensity and circular polarization. 
The Fe~{\sc i} line is more Zeeman-sensitive, since it blends two transitions with effective Landé factors of 1.25 and 1.37, whereas Sr~{\sc i} has a value of 1.0 \citep[based on LS-coupling;][]{LandiDeglInnocenti1982}. 
Thus, if magnetic fields with significant vertical component were present, Fe~{\sc i} should exhibit stronger polarization: circular at disk center, and both circular and linear toward the limb. 
While the circular polarization in Fe~{\sc i} closely mirrors that in Sr~{\sc i}, its linear polarization remains weak, pointing to a different dominant physical origin for the stronger $P_\mathrm{L}$ signals in Sr~{\sc i}--namely, scattering polarization.
The strongest total linear polarization structures in Fe~{\sc i} are spatially more localized than those in Sr~{\sc i}. 
Whether these Fe~{\sc i} signals originate from Zeeman-induced polarization due to transverse magnetic fields or are instead due to residual instrumental systematics remains difficult to quantify. 
However, regardless of the cause, the much stronger and more extended $P_\mathrm{L}$ signals in Sr~{\sc i} strongly suggest that scattering is the dominant process generating the linear polarization in this line.

The low circular polarization amplitudes (typically $<1$\%) in both lines, as seen in the center panels of Fig.\ref{fig:maps}, further support the absence of significant LOS magnetic fields in the observed region. 
The similarity of $V/I$ and intensity maps between Fe~{\sc i} and Sr~{\sc i} indicates that both lines form at comparable heights and trace similar LOS magnetic structures.

The intensity structures appear finer along the scanning direction than along the slit. 
This asymmetry is most likely due to geometric foreshortening at $\mu=0.74$.
In this observation, the slit orientation tends to be parallel to the nearest solar limb (see Fig.~\ref{fig:maps}), so that granules  appear compressed in the scanning direction. 
Although atmospheric seeing can introduce distortions, such effects are typically random in time and space and do not produce stable, coherent structures across multiple scan positions. 
The persistence and spatial coherence of the observed features suggest that they are not caused by seeing-induced fluctuations.
We therefore interpret the observed structuring as genuine solar granulation, consistent with known sub-arcsecond-scale intensity patterns in the photosphere near the line's formation heights.

\section{Conclusions}

We have presented the first direct, two-dimensional observations of sub-arcsecond spatial structure in the scattering polarization of the Sr~{\sc i}~4607\,\AA\ line, obtained in quiet-Sun regions near disk center at $\mu=0.74$ with ViSP at DKIST.

The observed fine-scale patterns in the total linear polarization are well above the noise level and show spatial coherence along both the slit and scan directions. These properties, together with the absence of comparable structures in the simultaneously observed Fe~{\sc i} line, indicate that the detected signals in Sr~{\sc i} are of solar origin and dominated by scattering polarization rather than the Zeeman effect. 
If the Zeeman effect were the primary contributor, the  Fe~{\sc i} line, being more sensitive to the Zeeman effect than the Sr~{\sc i} line, should display stronger linear polarization, which is not observed.

The spatial scales of the detected structures are consistent with sub-granular patterns predicted by radiative transfer modeling of the scattering polarization in this line \citep{TrujilloBueno2007,DelPinoAleman2018,DelPinoAleman2021}, although a quantitative comparison will be the subject of future work. 
While our analysis includes careful checks for instrumental effects, we cannot entirely exclude a minor contribution from residual artifacts, particularly at the lowest signal levels.

This first direct mapping of sub-arcsecond scattering polarization demonstrates the diagnostic potential of high-resolution spectropolarimetry with ViSP and DKIST. 
Distinguishing the scattering polarization distribution relative to granules and intergranules is of particular scientific interest, but at the observed LOS, which was not too close to the disk center in order to have a sufficient signal-to-noise ratio, such a separation is not straightforward. 
A forthcoming study will address this analysis in detail.
These observations provide a new benchmark for testing and refining models of small-scale magnetism and scattering polarization in the solar photosphere, paving the way for future high-resolution studies of quiet-Sun magnetism.

\section*{Acknowledgments}
F.Z. and L.B. acknowledge funding from the Swiss National Science Foundation under grant numbers PZ00P2\_215963 and 200021\_231308, respectively. T.P.A.'s participation in the publication is part of the Project RYC2021-034006-I, funded by MICIN/AEI/10.13039/501100011033, and the European Union “NextGenerationEU”/RTRP.
T.P.A. and J.T.B. acknowledge support from the Agencia Estatal de Investigaci\'on del Ministerio de Ciencia, Innovación y Universidades (MCIU/AEI) under grant
``Polarimetric Inference of Magnetic Fields'' and the European Regional Development Fund (ERDF) with reference PID2022-136563NB-I00/10.13039/501100011033.
E.A.B. acknowledges financial support from the European Research Council (ERC) through the Synergy grant No. 810218 (``The Whole Sun'' ERC-2018-SyG). 
Work by R.C. was supported by the National Center for Atmospheric Research, which is a major facility sponsored by the National Science Foundation (NSF) under Cooperative Agreement No.~1852977.
This research has made use of NASA's Astrophysics Data System Bibliographic Services. 
The research reported herein is based in part on data collected with the Daniel K. Inouye Solar Telescope (DKIST), a facility of the National Solar Observatory (NSO). The NSO is managed by the Association of Universities for Research in Astronomy, Inc., and funded by the National Science Foundation. Any opinions, findings, and conclusions or recommendations expressed in this publication are those of the authors and do not necessarily reflect the views of the National Science Foundation or the Association of Universities for Research in Astronomy, Inc. DKIST is located on land of spiritual and cultural significance to Native Hawaiian people. The use of this important site to further scientific knowledge is done with appreciation and respect. The observational DKIST data used during this research is openly available from the \href{https://dkist.data.nso.edu}{ DKIST Data Center Archive} under the proposal identifier pid$\_2\_70$.

\bibliographystyle{bibtex/aa}
\bibliography{bibtex/bib}
\end{document}